\documentclass[showpacs,aps,prl,twocolumn,floatfix,superscriptaddress,nofootinbib]{revtex4}
\usepackage{graphicx}
\usepackage{bm}
\usepackage{amsmath}

\newcommand{\be}{\begin{equation}}
\newcommand{\e}{\end{equation}}
\newcommand{\beml}{\begin{subequations}}
\newcommand{\eml}{\end{subequations}}
\newcommand{\beq}{\begin{eqnarray}}
\newcommand{\eq}{\end{eqnarray}}
\newcommand{\ba}{\begin{array}}
\newcommand{\ea}{\end{array}}
\newcommand{\lt}{\left}
\newcommand{\rt}{\right}

\newcommand{\la}{\langle}
\newcommand{\ra}{\rangle}

\newcommand{\re}{\,{\rm Re}\,}
\newcommand{\ep}{\boldsymbol{\varepsilon}}

\newcommand{\Du}{\textbf{D}^{\dagger}}
\newcommand{\Dd}{\textbf{D}}
\newcommand{\vk}{\textbf{k}}

\newcommand{\bra}[1]{\left|#1\right\rangle}
\newcommand{\ket}[1]{\left\langle#1\right|}

\begin{document}

\date{November 19, 2004}

\title{Coherent inelastic backscattering of intense laser light 
by cold atoms}

\author{V.~Shatokhin}
\affiliation{B.~I.~Stepanov Institute of Physics, National Academy of Sciences
  of Belarus, Skaryna Ave. 70, BY-220072  Minsk, Belarus}
\affiliation{Max-Planck-Institut f\"ur Physik komplexer Systeme,
N\"othnitzer Str. 38, D-01187 Dresden, Germany}
\author{C.~A.~M\"uller}
\affiliation{Physikalisches Institut, Universit\"at Bayreuth, D-95440
Bayreuth, Germany}
\author{A.~Buchleitner}
\affiliation{Max-Planck-Institut f\"ur Physik komplexer Systeme,
N\"othnitzer Str. 38, D-01187 Dresden, Germany}

\begin{abstract}
We present a nonperturbative treatment of coherent backscattering of intense laser
light from cold atoms, 
and predict a
nonvanishing backscattering signal even at very large intensities, due to the
constructive (self-)interference of inelastically scattered photons.
\end{abstract}

\maketitle

When a plane wave of arbitrary nature is incident upon a 
disordered 
medium of scatterers, the backscattered intensity 
is an interference pattern of all coherent partial amplitudes
containing detailed information on the sample configuration. Under an ensemble
average, interference between uncorrelated amplitudes is washed out,
except for a small angular range around exact backscattering,  
where the average intensity  
may exhibit a narrow
peak. This peak results from constructive interference between 
multiple scattering probability amplitudes
counterpropagating along direct and reversed paths
\cite{local,sheng}. 
This phenomenon 
is called 
Coherent Backscattering (CBS) 
and was for the first time 
demonstrated with 
samples of polystyrene particles \cite{clasCBS}. 
The CBS enhancement factor $\alpha$,  
the ratio of the total intensity at exact backscattering to the
background intensity, 
measures the coherence of counterpropagating amplitudes responsible
for localization effects. 
Recently, CBS
of light has been imported
to the quantum realm 
with clouds of cold 
atoms \cite{labeyrie99,katalunga03,chaneliere03}. 
An important leitmotiv of these 
studies is the robustness of the underlying interference effect with respect
to fundamental quantum mechanical dephasing mechanisms, such as spin-flip (of
the incoming radiation, which carries a polarization degree of freedom)
\cite{mueller02} or inelastic scattering. This has important repercussions for
the transition from weak to strong (in Anderson's sense) localization of light
in disordered atomic samples \cite{gora}, and
also for potential technological applications such as random lasers \cite{cao}. 
While in the regime of weak, perturbative atom-field coupling, 
the partial destruction of CBS due to spin-flip like processes --
induced by the multiple degeneracy of the atomic transition driven by the
incident radiation -- has been
demonstrated experimentally and analysed theoretically in quite some detail,
experiments and
theory only now start to probe the strong coupling limit, where 
inelastic photon-atom scattering processes prevail. First experimental results
on Sr (driving the $^1\!S_0\rightarrow ^{1}\!\!\!P_1$
transition with its nondegenerate ground state, hence in the absence of
spin-flip) \cite{chaneliere03} indeed 
demonstrate the reduction of the CBS enhancement factor with increasing
intensity of the injected field, for values
$s=\Omega^2/2(\Delta^2+\gamma^2)<1$ of the
atomic saturation parameter (where $\Omega$ is the Rabi frequency induced by
the driving, $\gamma$ half the 
spontaneous decay rate of the excited atomic level, and $\Delta$ the detuning
of the injected laser frequency from the exact atomic resonance). 
A first scattering theoretical treatment
\cite{wellens} identified the origin of such suppression in the availability
of which-path information through inelastically scattered photons: reversed
paths can be distinguished  by the detection of photons of different
frequency. However, this treatment, still perturbative in the field
intensity, cannot address the limit of large saturation parameters $s\geq 1$
-- with an emerging Mollow triplet \cite{mollow69} in the single atom resonance
fluorescence -- 
and, in particular, makes no prediction on the crossover from dominantly
elastic to essentially inelastic CBS, nor on the CBS enhancement factor in the
deep inelastic limit. In the present Letter, we enter this regime, starting
from a general master equation which allows for a nonperturbative treatment of
the atom-field coupling. As we will see, even inelastically scattered photons
give rise to a nonvanishing CBS signal.

We start out from the elementary toy model of CBS -- a laser field scattering
off two atoms with labels $1$ and $2$, placed at a fixed \cite{note} 
distance $r_{12}\gg
\lambda = 2\pi/k_L$, with ${\bf k}_L$ the wave vector of the incident field. It is
known from the perturbative treatment of CBS that double scattering (on two
atoms) provides the leading contribution to the CBS signal, since this is the
lowest order process which gives rise to time reversed scattering amplitudes
which can interfere constructively. We expect that this 
scenario also allows for a qualitative assessment of the nonlinear atomic
response in the regime of high laser intensities, whilst propagation effects
in the bulk of the scattering medium are certainly beyond reach of this
model. Disorder will be mimicked by a suitable average over the atomic
positions.
Our model also neglects the acceleration of atoms out of
resonance, which 
certainly becomes important
at very high intensities, but can be 
experimentally compensated for by shortening the CBS probe 
duration, as realized in  
\cite{chaneliere03}. 
We focus exclusively on the
photon coupling to the internal atomic degrees of freedom. 

\begin{figure}
\includegraphics[width=6cm]{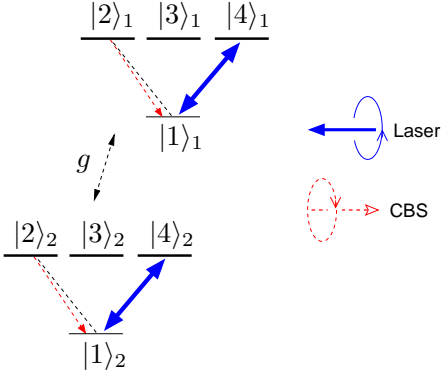}
\caption{(Color online)
Elementary configuration for coherent backscattering (CBS) of intense light
(thick arrows) by two isotropic dipolar transitions in the helicity preserving
polarization channel (dashed arrows).
The sublevels
$\bra 1$ and $\bra 3$ of both atoms have magnetic quantum number
$m=0$; sublevels $\bra 2$ and $\bra 4$ correspond to $m=-1$ and $m=1$, 
respectively.  
}
\label{fig:scat}
\end{figure}

With these premises, the average backscattered intensity 
can be derived
from the correlation functions of the atomic dipoles which emit the detected
signal \cite{agarwal,cohen_tannoudji}. We specialize to the scenario of the Sr
experiments,  
with a
nondegenerate atomic dipole transition $J_g=0\rightarrow J_e=1$,
driven by 
laser photons with right circular polarization on the sublevels $\bra
1\rightarrow \bra 4$ -- see Fig.~\ref{fig:scat}. The scattered light is
detected in the helicity preserving channel (i.e., of
photons which originate from the $\bra 2\rightarrow\bra 1$ transition), where
single scattering is absent. 
%%%%%%%%%%%%%%%%%%%%%%%%%%%%%%%%%%%%%%%%%%%%%%%%%%%
%%%%%%%%%%%%%%%%%%%%%%%%%%%%%%%%%%%%%%%%%%%%%%%%%%%%%%%%%%
Thus we obtain, up to an irrelevant prefactor,  
the expectation value for the stationary intensity scattered into the direction ${\bf k}$ %%@
close to the backward direction $-{\bf k}_L$, 
\be
\la I\ra_{\rm ss}=\la\sigma^1_{22}\ra_{\rm
  ss}+\la\sigma^2_{22}\ra_{\rm ss}
+2\re(\la\sigma^1_{21}\sigma^2_{12}\ra_{\rm ss} e^{i{\bf k}\cdot{\bf
r}_{12}})\, ,
\label{corr_f}
\e
where $\sigma^\alpha_{kl}\equiv \bra k_\alpha\ket l_\alpha$, for atom $\alpha$.
The steady-state values for correlation
functions of the form $\la \sigma^\alpha_{ij}\ra_{\rm ss}$ or
$\la\sigma^\alpha_{ij}\sigma^\beta_{kl}\ra_{\rm ss}$
can be found from the master equation  \cite{lehmberg} 
\be
\dot Q =\sum_{\alpha=1}^2{\cal L}_\alpha Q+\sum_{\alpha\neq \beta=1}
^2{\cal L}_{\alpha\beta}Q, 
\label{meq}
\e    
where the Liouvillians ${\cal L}_\alpha$ and ${\cal L}_{\alpha\beta}$
govern the evolution of 
an arbitrary atomic operator $Q$ for independent and dipole-dipole
interacting (through the exchange of one or several photons) 
atoms, respectively. 
$Q$ stands for 
an operator from the complete set of operators acting on a tensor product 
of Hilbert spaces
of individual atoms. For our choice of the atomic structure 
shown in Fig.~\ref{fig:scat},
\be
Q\in\underbrace{\{\sigma^1_{11},\cdots,\sigma^1_{44}\}\otimes\{\sigma^2_{11},\cdots,\sigma^2_
{44}\}
}_{256 \; {\rm operators}}. 
\label{Q_2_at}
\e 
The explicit form of the interaction-picture Liouvillians ${\cal L}_\alpha$
and ${\cal 
L}_{\alpha\beta}$ derived in the  
standard dipole, rotating-wave, and Born-Markov
approximations can be shown to read:
\begin{widetext}
\beq
{\cal L}_\alpha Q & = & -i\Delta[\Du_\alpha\cdot\Dd_\alpha,Q]
-\frac{i}{2}[\Omega_\alpha(\Du_\alpha\cdot\ep_L)+\Omega^*_\alpha
(\Dd_\alpha\cdot\ep_L^*),Q]
+\gamma\lt(\Du_\alpha\cdot[Q,\Dd_\alpha]+[\Du_\alpha,Q]\cdot\Dd_\alpha\rt),\\
{\cal L}_{\alpha\beta}Q&=&\Du_\alpha\cdot\overleftrightarrow{\bf
    T}(g,{\bf\hat
n})\cdot[Q,\Dd_\beta]+[\Du_\beta,Q]\cdot\overleftrightarrow{\bf  
T}^*(g,{\bf\hat n})\cdot\Dd_\alpha\, ,
\label{Liouvillians}
\eq
\end{widetext}
where $\Delta=\omega_L-\omega_0$ is the detuning,
$\Omega_\alpha=\Omega e^{i\vk_L\cdot{\bf r}_\alpha}$ is the atomic 
(coordinate-dependent) Rabi frequency, and $\ep_L$ fixes 
the polarization of the laser field.
\be
\Dd_\alpha=-\ep_{-1}\sigma^\alpha_{12}+\ep_0\sigma^\alpha_{13}-\ep_{+1}\sigma^\alpha_{14}   
\e  
is the lowering dipole operator of atom $\alpha$, with $\ep_{\pm 1},\;\ep_0$
the 
unit vectors of the spherical basis.
The radiative dipole-dipole
interaction due to exchange of photons between the atoms is described
by the tensor 
$
\overleftrightarrow{\bf T}(g,{\bf\hat n})=\gamma g
\overleftrightarrow{\boldsymbol{\Delta}}$,  
where $\overleftrightarrow{\boldsymbol{\Delta}}=\overleftrightarrow
{\openone}-{\bf \hat{n}\hat{n}}$ is the projector on the 
plane 
defined by the vector $\bf\hat{n}$ connecting atom $1$ and
$2$, and 
\be
g =i3 \frac{e^{ik_0 r_{12}}}{2k_0 r_{12}},
\label{g}
\e
where $k_0=\omega_0/c$, is the small coupling constant
$|g|\ll 1$  in the far-field 
limit $k_0r_{12}\gg 1$, where we neglect near-field interaction
terms of order $1/(k_0r_{12})^2$ and $1/(k_0r_{12})^3$ 
(which, at higher
atomic densities, could also be retained in our formalism). 

Transforming the operator equation (\ref{meq}) to a system of 
$255$ linear
coupled differential equations 
for the
atomic correlation functions, we can solve for 
the physical quantities which enter the expression (\ref{corr_f}) for the
detected intensity. 
In doing so, we furthermore take advantage of the far field limit
$k_0r_{12}\gg 1$, and expand the correlation  
functions up to second order, ${\la(\dots)\ra}_{\rm ss}^{[2]}$, 
in the dipole-dipole coupling 
constant (\ref{g}). The double scattering contribution to the CBS signal,
detected in the helicity preserving channel, is
then precisely given by terms proportional to $g^2$, since it stems from the 
exchange of two photons between the atoms, along a `direct' and its `reversed'
path. 
Finally, the CBS signal is obtained after  
an elementary configuration average $\la\dots\ra_{\rm conf.}$ defined through the
following 
twofold procedure: (i) isotropic averaging of the relative orientation
${\bf \hat r}_{12}$ over the unit sphere; (ii) uniform averaging of
the 
distance $r_{12}$ over an interval of
the order of $\lambda$, around a mean value given by the mean free path.
After this simple procedure all terms 
relevant for the calculation of the CBS enhancement factor survive,
whereas all the irrelevant terms vanish.  

We thus arrive at our final expression for 
the total second-order intensity
\be
I_{\rm ss}^{\rm tot\,[2]}(\theta)=L^{\rm tot}+C^{\rm tot}(\theta),
\label{d_sc}
\e
a sum of the total ladder (or background),
$L^{\rm tot}$, and total crossed (or interference) term $C^{\rm
  tot}(\theta)$, with $\theta$ the observation angle of the scattered
intensity with respect to the backward direction.  
In terms of the second order atomic correlation functions, $L^{\rm tot}$ and
$C^{\rm 
  tot}(\theta)$ are given by
\beq
L^{\rm tot}&=&\lt\la\la\sigma_{22}^1\ra_{\rm ss}^{[2]}+\la\sigma_{22}^2\ra_{\rm 
ss}^{[2]}\rt\ra_{\rm conf.},\label{l1}\\
C^{\rm tot}(\theta)&=&2\re\lt\la\la\sigma_{21}^1\sigma_{12}^2\ra_{\rm
ss}^{[2]}e^{i{\bf  
k}\cdot{\bf r}_{12}}\rt\ra_{\rm conf.}.
\label{c1}
\eq
Therefrom we deduce 
the main quantifier
of CBS, the enhancement factor 
\be
\alpha=\frac{L^{\rm tot}+C^{\rm tot}(0)}{L^{\rm tot}}.
\label{enh_f}
\e

\begin{figure}
\includegraphics[width=6cm]{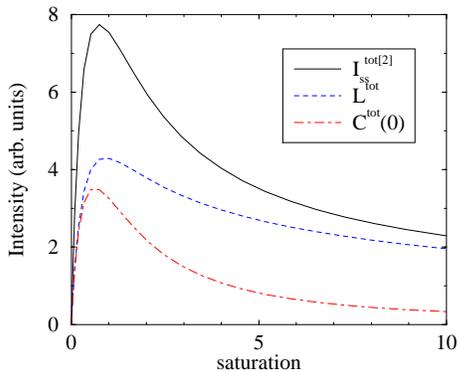}
\caption{(Color online)
Total intensities of the ladder, the crossed terms, and their sum
$I^{\rm tot\,[2]}_{\rm ss}$, in the helicity
preserving channel, as functions of the saturation parameter $s$
at resonance, $\Delta=0$.
}
\label{fig:int}
\end{figure}

Figure~\ref{fig:int} shows our results for the
total CBS intensity as well as its components $L^{\rm tot}$ and $C^{\rm
tot}(0)$, as a 
function of the saturation parameter, 
at exact resonance ($\Delta=0$). 
%%%%%%%%%%%%%%%%%%%%%%%%%%%%%%%%%  
%%%%%%%%%%%%%%%%%%%%%%%%%%%%%%%%%%%%%%%%%%%%%%%%%%%%%%%
The behavior of $I^{\rm tot\,[2]}_{\rm ss}$ shows
that the double scattering 
intensity behaves markedly different from that of an isolated atom. Whilst the  
scattering intensity from an isolated atom $I^{[0]}\propto s/(1+s)$ is known 
to
saturate for large $s$ \cite{cohen_tannoudji}, 
the double scattering intensity exhibits a maximum at $s\simeq 0.7$, 
followed by gradual decrease $\sim s^{-1}$ for large $s$: At high laser
intensities, more and more photons are scattered inelastically,  
and are therefore less likely to undergo resonant interaction with 
the second atom. 

%%%%%%%%%%%%%%%%%%%%%%%%%%%%%%%%%  
\begin{figure}
\includegraphics[width=6cm]{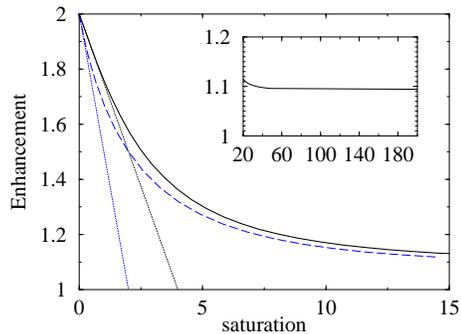}
\caption{(Color online)
Enhancement factor $\alpha$ in the helicity preserving channel, versus
saturation parameter $s$. Solid curve: on resonance ($\Delta =0$), dashed curve:
off resonance ($\Delta =\gamma$). Straight lines represent the
perturbative prediction $2-(1+\delta)s/4$ of \protect\cite{wellens}. 
Inset: The finite enhancement $\alpha_{\infty}\simeq 1.09$ signals 
residual photon (self-)interference, even in the deep
inelastic regime. }
\label{fig:enh}
\end{figure}
%%%%%%%%%%%%%%%%%%%%%%%%%%%%%%%%%%%%%%%%%%%%%%%%%%%%%%%%%%%%
The enhancement factor $\alpha(s)$ follows directly from the above
quantities.
As shown in Fig.~\ref{fig:enh}, 
it decays monotonously from its weak field limit $\alpha(0)=2$.
In qualitative agreement with the experiment \cite{chaneliere03}, this
decay is faster for finite detuning $\Delta =\gamma$. 
For small values of $s$, $\alpha$ is well approximated by the linear 
decay $2-(1+\delta)s/4$, with $\delta=(\Delta/\gamma)^2$,
derived within the scattering picture \cite{wellens}.
For large values of $s$ (inset), however, $\alpha$ saturates at a 
value $\alpha_{\infty}\simeq 1.09$ strictly larger than unity, whilst 
one would expect vanishing contrast (i.e., $\alpha =1$) for scattering from
two independent atoms \cite{kochan95}. Hence,
the (self-)interference of inelastically scattered photons unambiguously 
contributes to the crossed term $C^{\rm tot}(0)$. Note that this observation 
bears some similarity to
CBS with photons from degenerate Raman transitions, which were shown to
yield an important contribution to the CBS contrast, even in the limit of
infinite ground state degeneracy \cite{mueller01}, as well as to the residual 
CBS enhancement in optically active media at high magnetic fields
\cite{martinez}.  

In contrast, elastically scattered photons remain perfectly coherent, and
contribute to the CBS intensity with a constrast two, for any $s$. To see
this, we just need to extract the purely elastic component of the signal from
the total yield in eq. (\ref{corr_f}). Since the detected intensity 
$\la I\ra_{\rm ss}$ is nothing but the autocorrelation function of
the source field amplitudes radiated by the atomic dipoles, its elastic part 
$\la I\ra^{\rm el}_{\rm ss}$ is generated by the classical dipoles
induced by the injected radiation -- this is by their average,
nonfluctuating parts $\la\sigma^\alpha_{i\ne j}\ra_{\rm ss}$
\cite{cohen_tannoudji}.  
Hence, $\la I\ra^{\rm el}_{\rm ss}$ is given by the product of the expectation values of
the atomic dipoles: 
% rather than by the expectation value of their product
%(which immediately leads to eq.~(\ref{corr_f})), what gives
\be
\la I\ra^{\rm el}_{\rm ss}=|\la\sigma^1_{21}\ra_{\rm
  ss}|^2+|\la\sigma^2_{21}\ra_{\rm ss}|^2
+2\re(\la\sigma^1_{21}\ra_{\rm ss}\la\sigma^2_{12}\ra_{\rm ss} e^{i{\bf k}\cdot{\bf
r}_{12}})\, .
\label{el_corr_f}
\e
A power series expansion of the right hand side of (\ref{el_corr_f}) to second
order in the coupling $g$ leaves only symmetrically factorized combinations of
the form 
$\langle\sigma_{21}^{\alpha}\rangle^{[1]}\langle\sigma_{12}^\beta\rangle^{[1]}$.
Asymmetric combinations, like 
$\langle\sigma_{21}^{\alpha}\rangle^{[2]}\langle\sigma_{12}^\beta\rangle^{[0]}$,
do not contribute to the signal since the $|1\ra\leftrightarrow|2\ra$
transitions are not 
laser-driven (see Fig.~\ref{fig:scat}), hence
$\langle\sigma_{12}^\beta\rangle^{[0]}$  
vanishes.
Evaluation of the correlation functions by symbolic calculus, together with
the configuration average described above, finally provides 
an analytic
expression for the elastic ladder and crossed terms:
\be
L^{\rm el}=C^{\rm el}(0)=24\pi|g|^2
\frac{1}{1+\delta}\frac{s}{(1+s)^4}\, .
\label{coh_comp}
\e

Expression (\ref{coh_comp}) shows that the elastic ladder and crossed terms 
are equal for any $s$,
as to be expected from
reciprocity arguments \cite{mueller01}. 
These elastic components decay like  $s^{-3}$ at large saturation, much
faster than the total intensities that decrease like $s^{-1}$ (cf.\
Fig.~\ref{fig:int}). This proves that the residual CBS enhancement 
$\alpha_\infty$ is entirely due to the (self-)interference of
inelastically scattered photons. 
Furthermore, expression (\ref{coh_comp}) %it 
shows that the  
elastic part of the double scattering intensity exhibits a maximum at
$s=1/3$, slightly below the departure of $\alpha(s)$ from the perturbative
prediction of \cite{wellens} in Fig.~\ref{fig:enh}. Consistently, 
an expansion of (\ref{coh_comp}) to second order in
$s$ reproduces the expression $L^{\rm el}=C^{\rm el}(0)\sim s-4s^2$ derived
in \cite{wellens}.
Note that the crossover to the nonlinear regime for double scattering occurs
at a value of $s$ three times smaller than for 
an isolated atom, where $I^{\rm
el[0]}\propto s/(1+s)^2$ exhibits a maximum at $s=1$. This has a transparent
interpretation, by virtue of factorizing 
eq.~(\ref{coh_comp}) into 
(i) the elastic
intensity $I^{{\rm el}[0]}$ scattered by the first
strongly driven atom, 
(ii) the 
total scattering cross section $\sigma^{\rm tot}\propto
1/(1+\delta)(1+s)$ of the second atom,
and (iii) the relative weight $I^{\rm el[0]}/I^{\rm tot[0]}=\sigma^{\rm
el}/\sigma^{\rm tot}=(\gamma^2+\Delta^2)/(\gamma^2+\Omega^2/2+\Delta^2)=
1/(1+s)$ \cite{cohen_tannoudji} 
of elastic processes therein.
Obviously, higher order scattering processes than
considered in 
our present 
contribution must unavoidably push the crossover value of $s$ to even smaller
values. 

In conclusion, we have presented the first study of coherent backscattering of
intense laser light from saturated dipole transitions. The CBS enhancement
decreases monotonously as a function of $s$, but, remarkably, 
coherence is -- partially -- preserved in the
deep inelastic limit of the two-atom response to intense laser radiation,
since also inelastically scattered photons can interfere with themselves,
along time-reversed paths. Consequently, CBS should also have an imprint on
the spectrum of the scattered radiation, as well as on its photocount
statistics, which are both directly accessible in the framework of our present
approach, as well as in laboratory experiments. 
Furthermore, let us note that our 
present results are also relevant in the somewhat different 
context of Young's double slit experiments 
with two atoms  \cite{kochan95}. In contrast to the forward
Young-type interference that necessarily decoheres for $s\to \infty$, since
the photon visits two different, uncoupled atoms,  a
\emph{backscattering} experiment, with appropriate 
polarization sensitive excitation and detection, 
must lead to a finite interference contrast. 

\begin{acknowledgments}
We would like to thank Dominique Delande, Beno{\^\i}t Gr\'emaud, Christian 
Miniatura, Mikhail Titov, and Thomas Wellens for stimulating discussions. 

\end{acknowledgments}


\begin{thebibliography}{99}
\bibitem{note}
At typical temperatures of the cloud of about a few mK \cite{labeyrie99},
the Doppler shift associated with the thermal motion is negligible, yet
the kinetic energy of an atom is much larger than the 
recoil energy, thus allowing a classical description 
of the atomic position \cite{mueller01}.
\end{thebibliography}
\end{document}